\documentclass[manuscript]{aastex}
\usepackage[colorlinks=True,citecolor=blue,breaklinks=True]{hyperref}
\usepackage{graphicx}
\usepackage{rotating} 

\shorttitle{Turbulent heating in the solar wind}
\shortauthors{Sasikumar Raja et al.}

\begin{document}

\title{Turbulent density fluctuations and proton heating rate in the solar wind from $9-20~R_{\odot}$}

\author{K. Sasikumar Raja\altaffilmark{1}}
\affil{Indian Institute of Science Education and Research, Pashan, Pune - 411 008, India}
\email{sasikumar@iiserpune.ac.in}

\author{Prasad Subramanian\altaffilmark{1}}
\affil{Indian Institute of Science Education and Research, Pashan, Pune - 411 008, India}

\author{R. Ramesh\altaffilmark{2}}
\affil{Indian Institute of Astrophysics, 2nd Block, Koramangala, Bangalore - 560 034, India}

\author{Angelos Vourlidas\altaffilmark{3,4}}
\affil{Applied Physics Laboratory, Johns Hopkins University, Laurel, Maryland, USA}

\author{Madhusudan Ingale\altaffilmark{5}}
\affil{Plot No. 2, Near RSS office, Bamb Colony, Jammer Road, Bhusaval - 425 201, India}

\altaffiltext{1}{Indian Institute of Science Education and Research, Pashan, Pune - 411 008, India} 
\altaffiltext{2}{Indian Institute of Astrophysics, 2nd Block, Koramangala, Bangalore - 560 034, India}
\altaffiltext{3}{Johns Hopkins Applied Physics Laboratory, Laurel, Maryland, USA}
\altaffiltext{4}{also at IASAARS, National Observatory of Athens, Athens, GR-15236, Greece}
\altaffiltext{5}{Plot No. 2, Near RSS office, Bamb Colony, Jammer Road, Bhusaval - 425 201, India}

\begin{abstract}
We obtain scatter-broadened images of the Crab nebula at 80 MHz as it transits through the inner solar wind in 
June 2016 and 2017. These images are anisotropic, with the major axis oriented perpendicular to the radially 
outward coronal magnetic field. Using these data, we deduce that the density modulation 
index ($\delta N_e/N_e$) caused by turbulent density fluctuations in the solar wind
ranges from 1.9 $\times 10^{-3}$ to 7.7 $\times 10^{-3}$ between 9 —- 20 $R_{\odot}$. We also find that the heating rate of solar wind protons
at these distances ranges from $2.2 \times 10^{-13}$ to $1.0 \times 10^{-11} ~\rm erg~cm^{-3}~s^{-1}$.
On two occasions, the line of sight intercepted a coronal streamer. We find that the presence of the streamer approximately doubles the thickness of the scattering screen.

\end{abstract}

\keywords{Sun: solar wind -- Sun: corona -- Sun: radio radiation -- Occultations -- turbulence -- scattering}

\section{Introduction}\label{intro}

The solar wind exhibits turbulent fluctuations in velocity, magnetic field, and density. 
Traditionally, researchers have attempted to understand this phenomenon within the framework of 
incompressible magnetohydrodynamic (MHD) turbulence (e.g., \citet{Gol1995}). 
However, density fluctuations are not explained in this framework, and remain a relative enigma despite 
noteworthy progress (e.g., \citet{Hna2005, Sha2010, Ban2014}). While most of the data used for solar wind turbulence studies 
are from in-situ measurements made by near-Earth spacecraft, density fluctuations 
can often been inferred via remote sensing observations, typically at radio wavelengths. Examples include 
angular broadening of point-like radio sources observed through the solar wind
\citep{Mac52, Hew63, Eri1964, Ble1972, Den1972, Sastry1974, Arm90, Ana94, Ram1999, Ram01, Ram2012, Kat2011, Mug2016, Sas2016}, interplanetary scintillations 
(IPS; \citet{Hew64, Coh69, Eke71, Ric90, Bis09, Man2000, Tok12, Tok16}),
spacecraft beacon scintillations \citep{Woo79}, interferometer phase scintillations using Very Long Baseline 
Interferometers (VLBI; \citet{Cro1972}),
spectral broadening using coherent spacecraft beacons \citep{Woo79} and radar echoes \citep{Har1983}.

A related problem is the issue of turbulent heating in the inner solar wind. 
It is well known that the expansion of the solar wind leads to adiabatic cooling, which is offset by some sort of 
heating process \citep{Ric1995, Mat1999}. The candidates for such extended heating range from resonant wave heating 
\citep{Cra2000,Hol2002} to reconnection events
(e.g., \citet{Car2004}). Some studies have attempted to link observations 
of density turbulence with kinetic Alfven waves that get resonantly damped on protons, consequently heating them \citep{Ing2015b, Cha2009}. 

In this paper, we investigate the 
characteristics of turbulent density fluctuations and associated solar wind heating rate from $9-20~R_{\odot}$ using the anisotropic 
angular broadening of radio observations of the Crab nebula from June 9 to 22 in 2016 and 2017. The Crab nebula passes 
close to the Sun on these days every year. Since its radiation passes through the foreground solar wind, these observations 
give us an opportunity to explore the manner in which its angular extent is broadened due to scattering off turbulent density 
fluctuations in the solar wind. Anisotropic scatter-broadening of background sources observed through the solar wind has hitherto 
been reported only for small 
elongations ($\approx 2-6 ~R_{\odot}$) e.g., \citep{Ana94,Arm90}. Imaging observations of the Crab nebula (e.g., 
\citet{Ble1972, Den1972}) offer us an opportunity to investigate this phenomenon for elongations 
$\gtrsim 10 R_{\odot}$. On 17 June 2016, 17 and 18 June 2017, a coronal streamer was present along the line of sight to the Crab nebula; 
this gives us an additional opportunity to study streamer characteristics. The Parker Solar Probe \citep{Fox2016} 
is expected to sample the solar wind as close as 10 $R_{\odot}$. In-situ measurements from the 
SWEAP instrument aboard the PSP can validate our findings regarding the density turbulence level and the proton heating rate.

The rest of the paper is organized as follows: in \S~2, we describe imaging observations of the Crab nebula made at Gauribidanur in June 2016 and 2017. 
The next section (\S~3) explains the methodology for obtaining the turbulence levels from these images. This includes a brief discussion 
of the structure function, some discussion of the inner scale of the density fluctuations, followed by the prescription we follow 
in computing the density fluctuations and solar wind heating rate at the inner scale. \S~4 summarizes our main results and conclusions.

\section{Observations: scatter-broadened images of the Crab nebula}\label{observations}

The radio data were obtained with the Gauribidanur RAdioheliograPH (GRAPH; \citet{Ram98,Ram11}) at 80 MHz during the local meridian transit of
the Crab nebula. The GRAPH is a T-shaped interferometer array with baselines ranging 
from $\approx 80$ to $\approx 2600$ meters. The angular resolution is $\approx$ 5 arcmin at 80 MHz, and the minimum detectable 
flux ($5 \sigma$ level) is $\approx 50$ Jy for 1 sec integration time and 1 MHz bandwidth. 
Cygnus A was used to calibrate the observations. Its flux density is $\approx 16296$ Jy at 80 MHz. 
The flux density of Crab nebula (when it is far from the Sun and is not therefore scatter-broadened by solar coronal turbulence) 
is $\approx 2015$ Jy at 80 MHz. We imaged the Crab nebula at different projected heliocentric distances
shown in column (3) of Table-\ref{tab:table-1} in the years 2016 and 2017.

\begin{figure*}[!ht]
\begin{minipage}{0.4\textwidth}
\centerline{\includegraphics[width=1.4\textwidth]{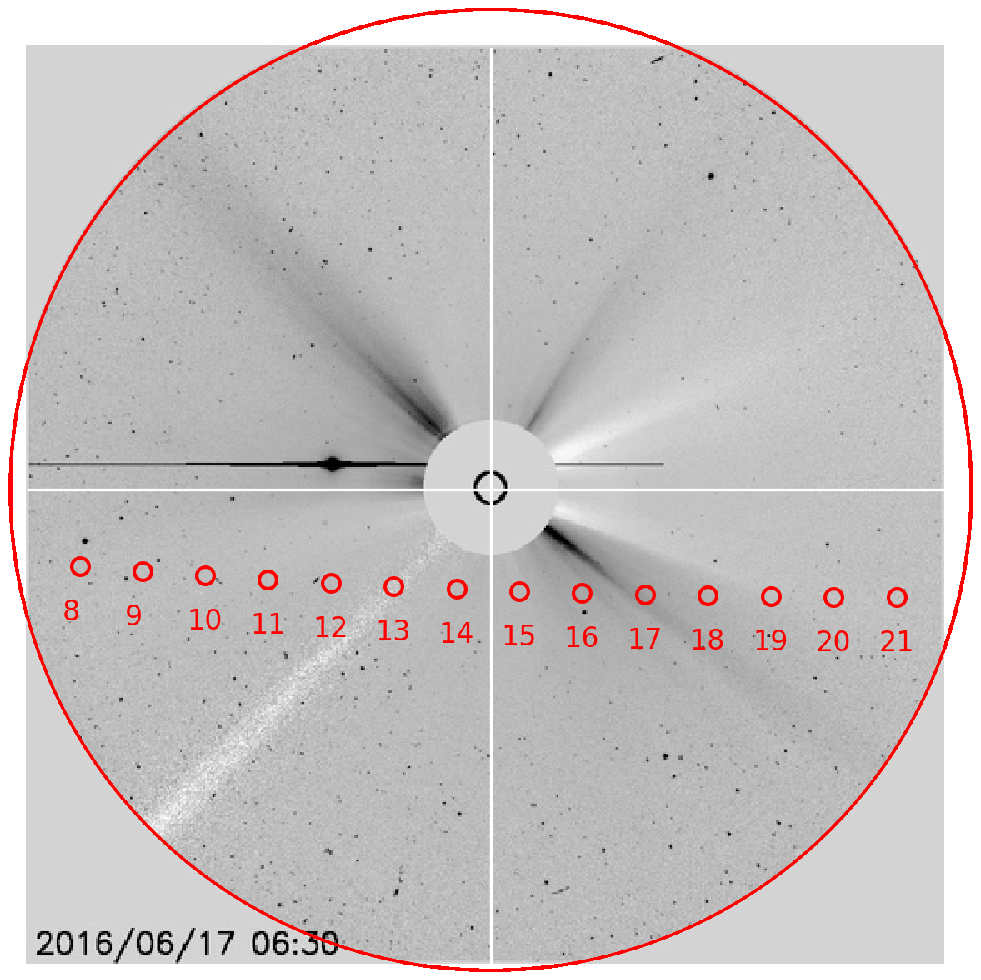}}
\end{minipage}
\begin{minipage}{0.6\textwidth}
\centering
\includegraphics[width=0.75\textwidth]{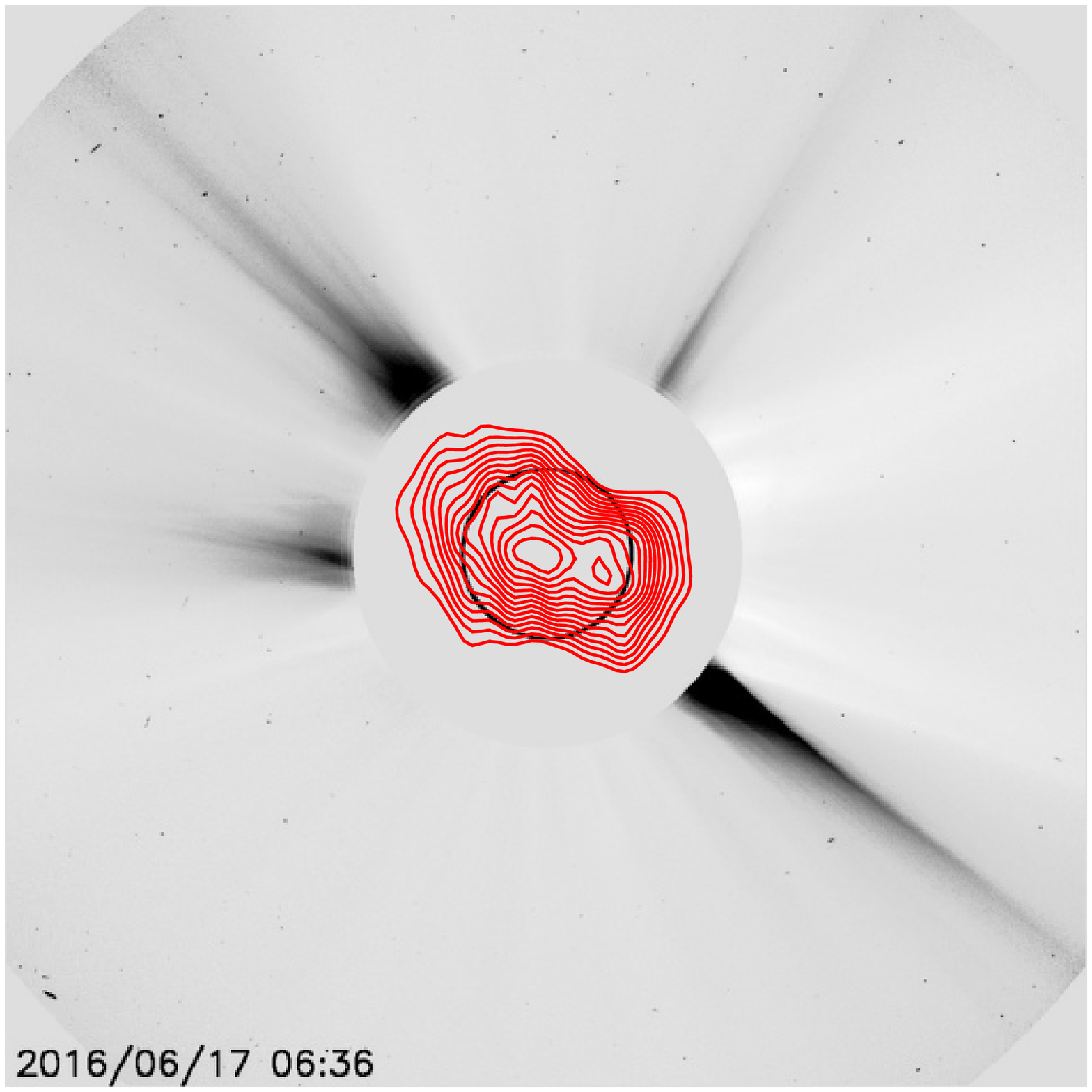}
\end{minipage}
\caption{(left) SOHO/LASCO C3 image of the solar corona (inverted grey scale image) observed on 17 June 2016 at 06:30 UT is shown. 
The innermost black circle indicates the solar disk (radius $=1~R_{\odot}$). 
The next concentric circle is the occulting disk of the coronagraph and its radius is $3.5~R_{\odot}$. 
The outermost circle marks a heliocentric distance of $30~R_{\odot}$. In both the images, the black features are coronal streamers. Solar north is up 
and east is to the left.
The small circles superposed on the image represent the location of the Crab nebula on 
different days during the period 8 June 2016 to 21 June 2016. 
Its closest approach to the Sun is on 14 June 2016 at a heliocentric distance of $\approx 5~R_{\odot}$.
The coronal streamer in the south-west quadrant occults the Crab nebula on 17 June 2016 
at a projected heliocentric distance $\approx 10.2~R_{\odot}$. 
The position angle (PA, measured counterclockwise from north) of the streamer is $\approx 235^{\circ}$.
(right) SOHO/LASCO C2 image of the solar corona (inverted grey scale) on 17 June 2016 at 06:36 UT is shown.
The red contours represent observations of the solar corona using the GRAPH at 80 MHz. The elongated radio contours
correspond to emission from the streamers in north-east and south-west quadrants.}
\label{fig:lasco}
\end{figure*}

We have used white light images of the solar corona obtained with the Large Angle and Spectrometric Coronagraph (LASCO)
onboard the SOlar and Heliospheric Observatory (SOHO) \citep{Bru95} for general context, and to identify features like coronal streamers.
Figure \ref{fig:lasco} shows the white light images of the solar corona obtained with the LASCO C3 (left) and C2 (right) coronagraphs 
on 17 June 2016. The black features in both inverted grey scale images are coronal streamers. 
The location of the Crab nebula between 8 and 21 June 2016 is marked by the red circles on the LASCO C3 images. 
On 17 June 2016, the Crab nebula was observed through a streamer in the south-west quadrant. The streamer was associated with 
an active region NOAA 12555 located at heliographic coordinates S09W71.
The contours superposed over the LASCO C2 image are from the GRAPH observations at 80 MHz showing radio emission 
from the streamers in north-east and south-west quadrants \citep{Ram2000}.

\begin{figure*}[!ht]
\centering
\begin{tabular}{cccc}
\includegraphics[width=.45\textwidth]{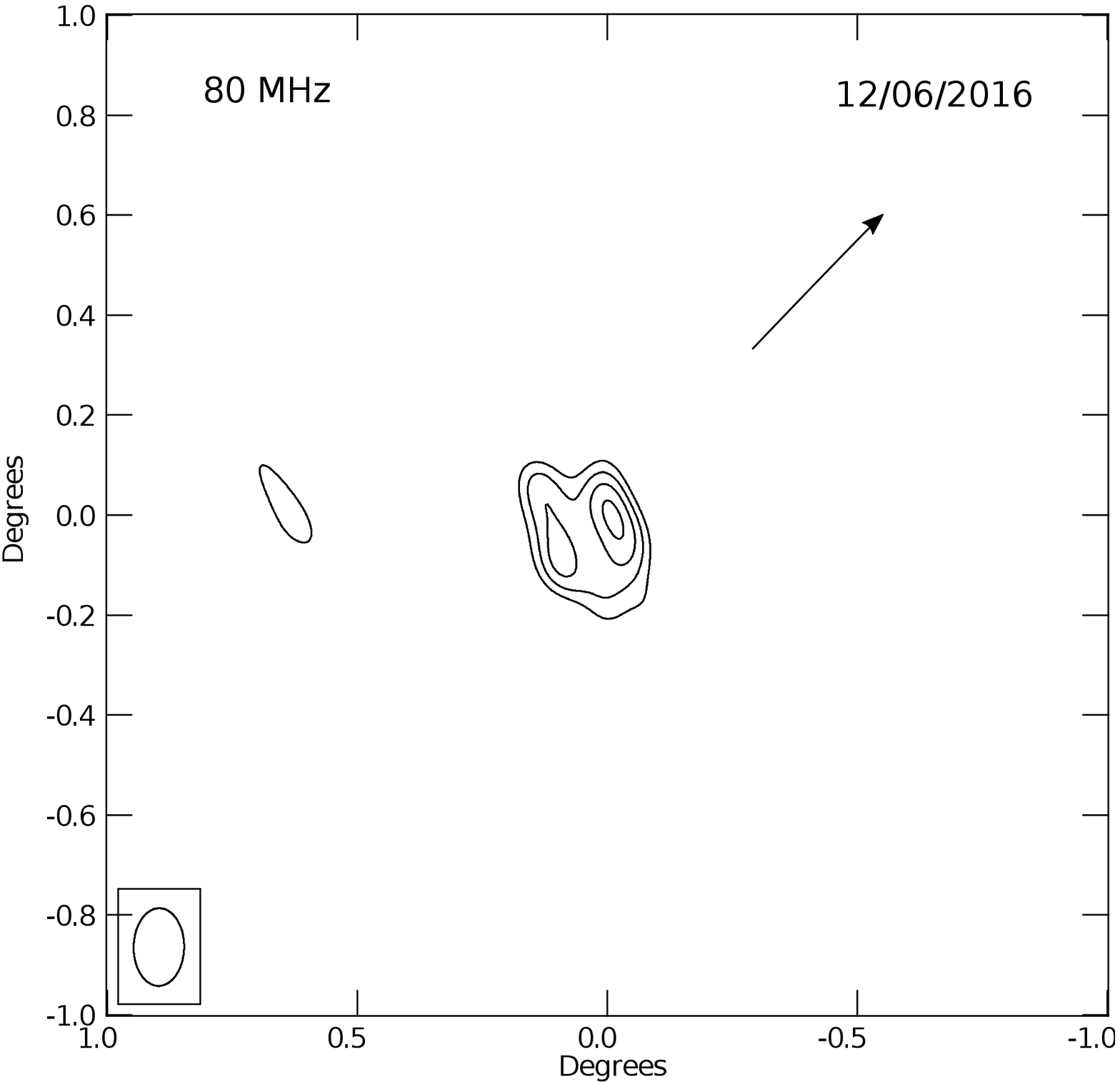} &
\includegraphics[width=.45\textwidth]{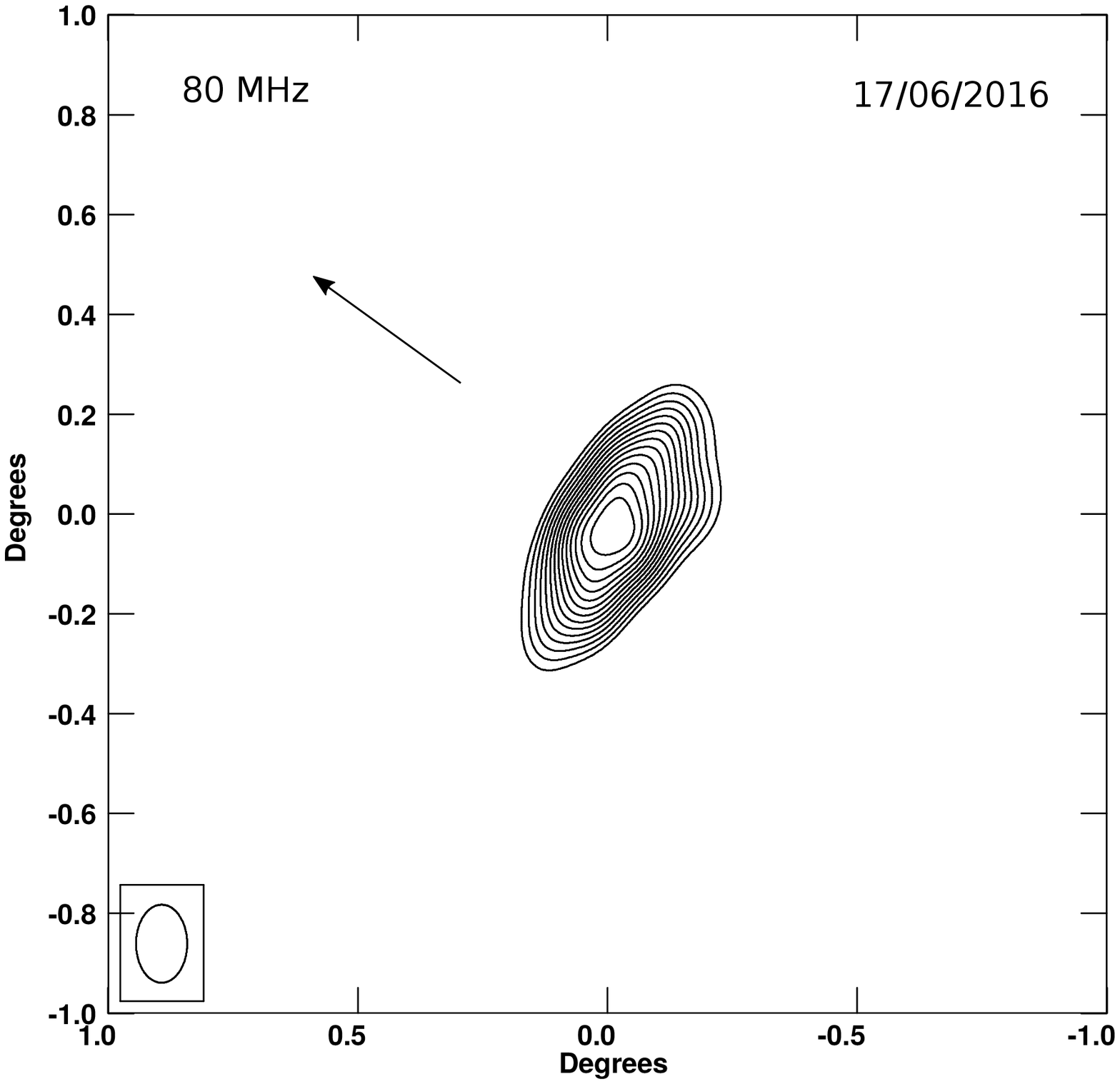} \\
\includegraphics[width=.45\textwidth]{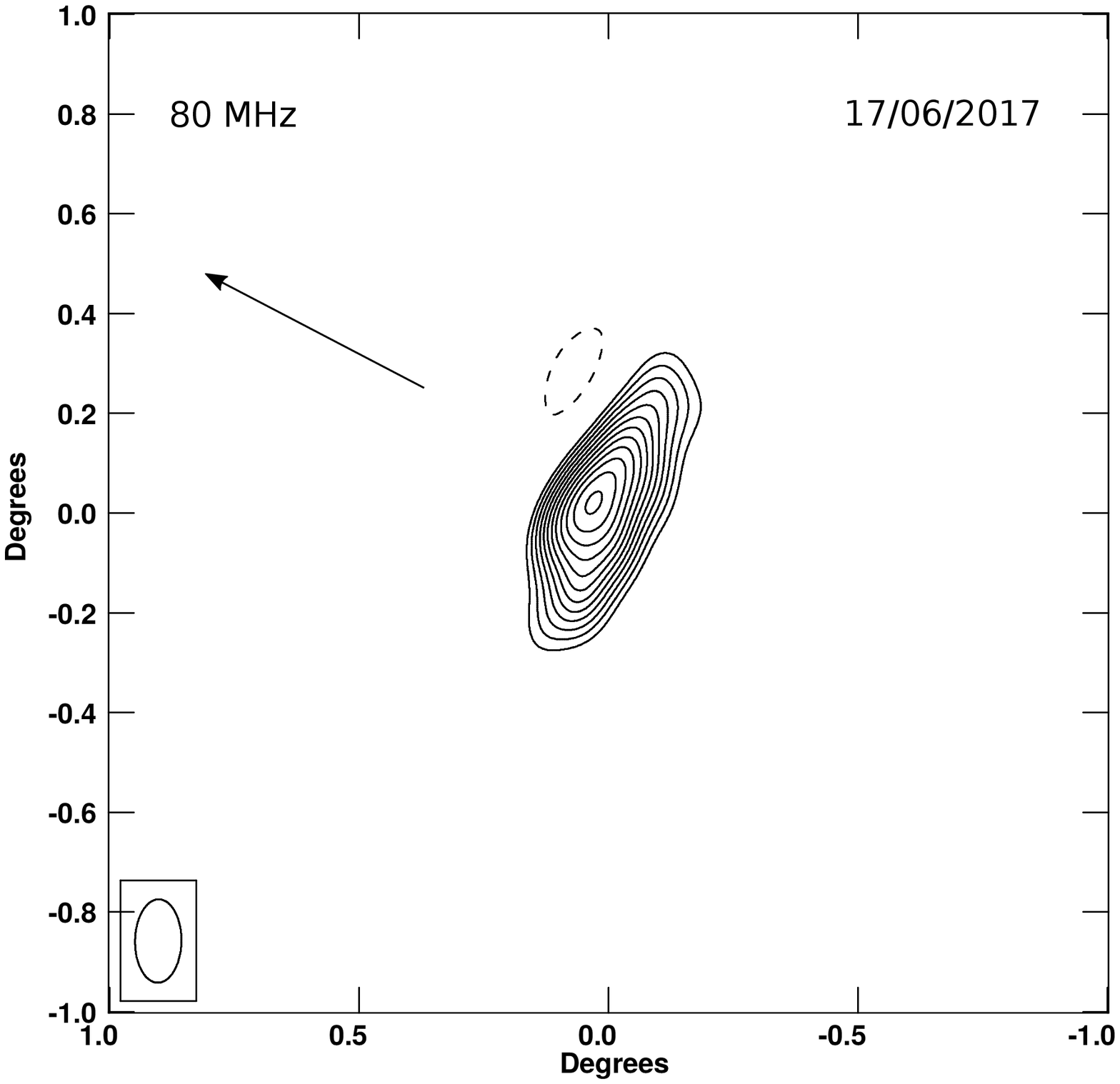} &
\includegraphics[width=.45\textwidth]{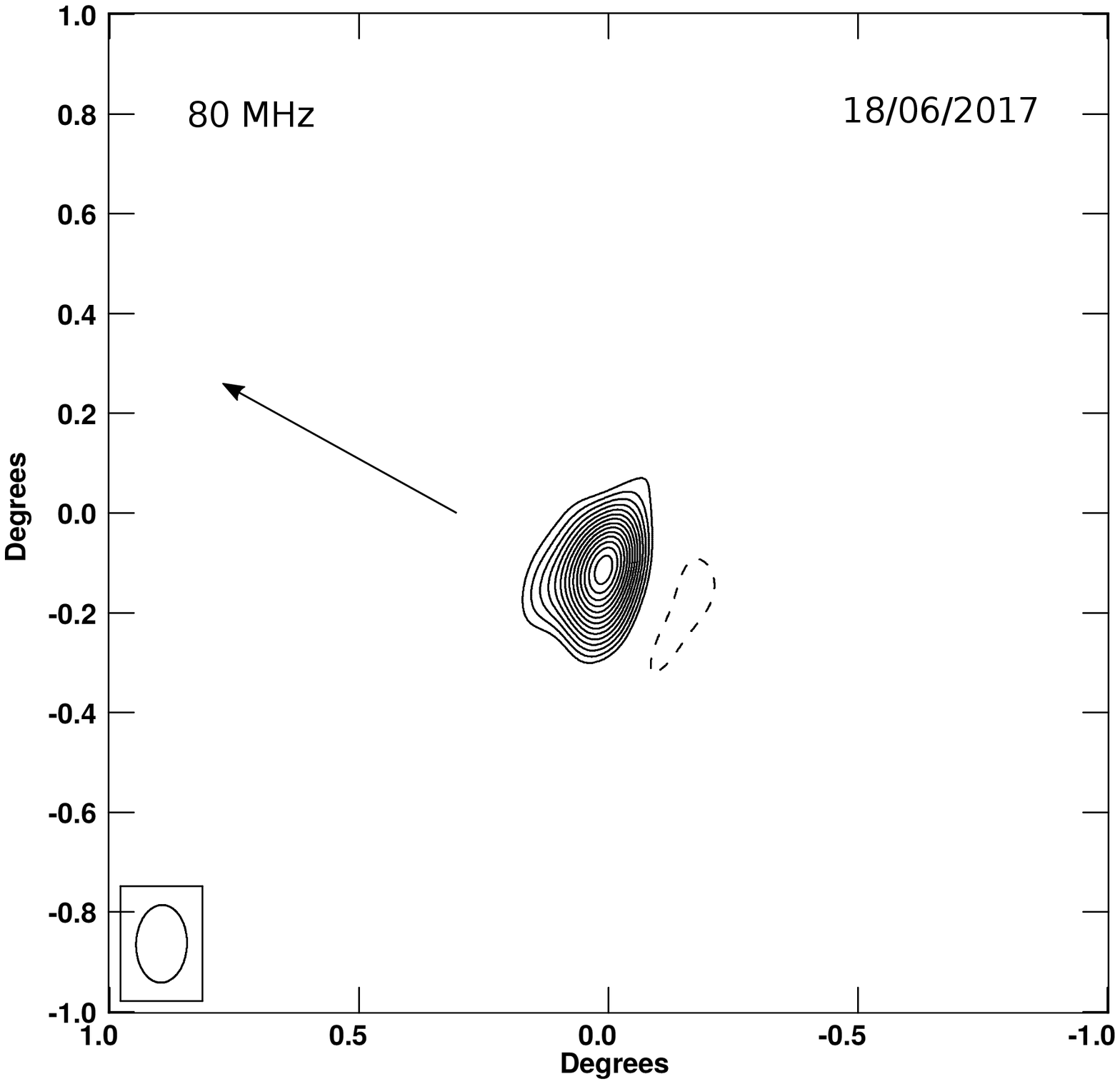}  \\ 
\end{tabular}
\caption{The image on 12 June 2016 shows the scatter broadened Crab nebula at a projected heliocentric distance of $10.18~ R_{\odot}$ during its ingress into the inner solar wind.
The images on 17 June 2016 (at $10.2~ R_{\odot}$), 17 ($9.41~ R_{\odot}$) and 18 June 2017 ($12.61~ R_{\odot}$) depict the scatter broadened Crab nebula 
observed through coronal streamers during its egress from the solar wind. The arrows depict the sunward direction on each day. The major axis of each 
image is perpendicular to the magnetic field lines, which are directed radially outward from the Sun.}
\label{fig:graph_images}
\end{figure*}

Some representative 80 MHz GRAPH images of the Crab nebula are shown in Figure \ref{fig:graph_images}. The image on 12 June 2016 was observed through 
the solar wind at $10.18~ R_{\odot}$ during ingress. The one on 17 June 2016 was observed at $10.20~ R_{\odot}$, while the one on 17 June 2017 at $9.41~ R_{\odot}$ 
and the one on 18 June 2017 at $12.61~ R_{\odot}$ during egress. The Crab nebula was occulted by a coronal streamer on 17 June 2016 and on 17 and 18 June 2017. 
These scatter-broadened images are markedly anisotropic. This aspect has been noted earlier, for the Crab nebula \citep{Ble1972, Den1972} as 
well as other sources \citep{Ana94,Arm90}. Note that the major axis of these images is always perpendicular to the heliocentric radial direction 
(which is typically assumed to be the magnetic field direction at these distances) - this is especially evident when the Crab is occulted by a streamer.
The parameters for all observations of the Crab nebula in 2016 and 2017 are tabulated in Table \ref{tab:table-1}.

\begin{figure}[!ht]
\centerline{\includegraphics[width=12cm]{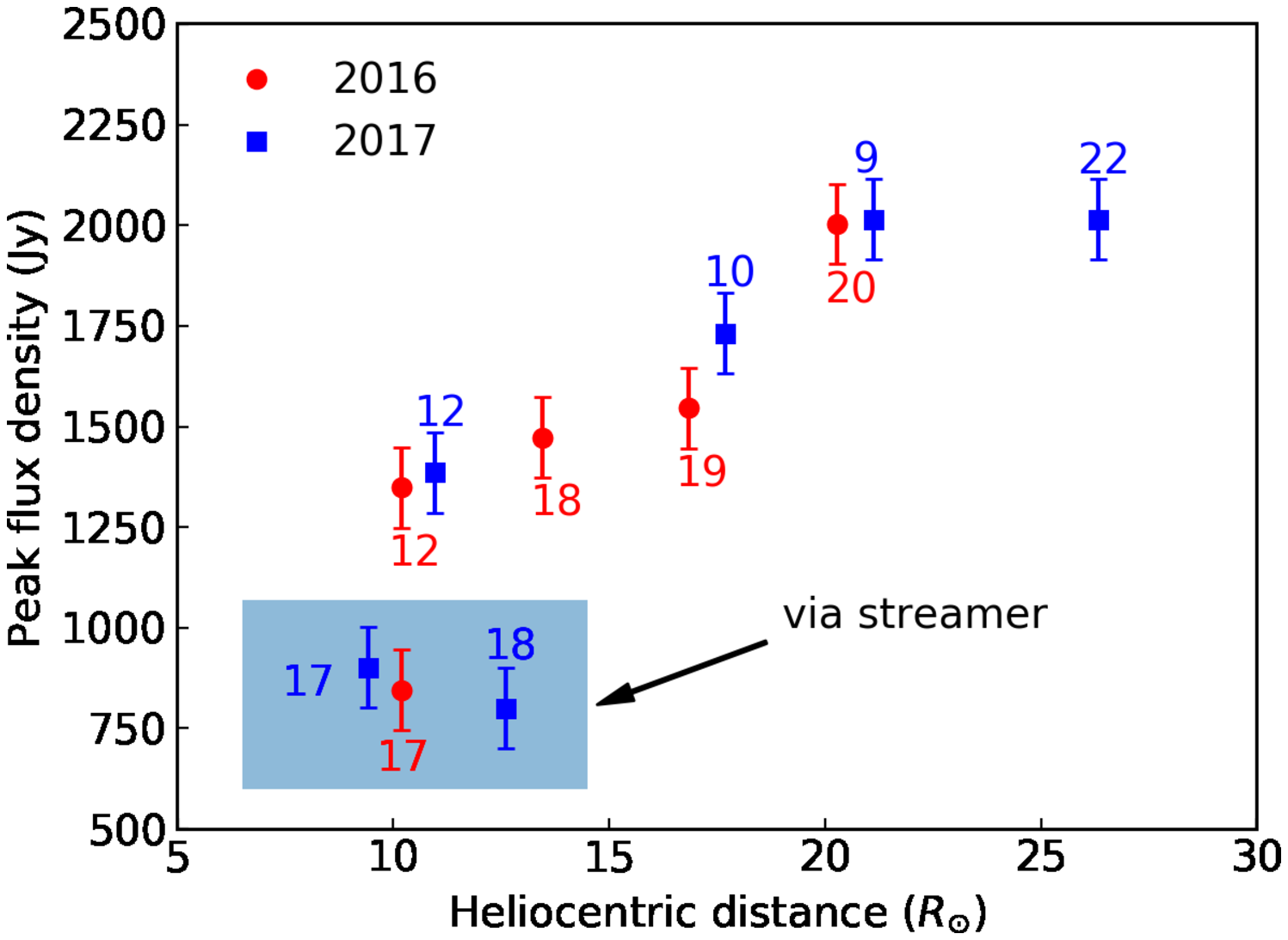}}
\caption{Peak flux density of the Crab nebula on different days of June 2016 (red circles)
and 2017 (blue squares). 
The red and blue data points shown 
in the shaded area indicate instances when the Crab nebula was observed through a streamer in 
2016 and 2017 respectively. 
} 
\label{fig:peak}
\end{figure}

Figure \ref{fig:peak} shows the observed peak flux density of the Crab nebula with respect to its projected heliocentric distance. 
The red circles and blue squares are for the 2016 and 2017 observations respectively.
Note that, in a given year the data points obtained during ingress and egress were plotted together with the (projected) heliocentric distance. 


The observations shown in the shaded region in Figure \ref{fig:peak} represent instances where the Crab nebula 
was occulted by a coronal streamer. Evidently, the peak flux density in these instances in considerably lower 
(as compared to the flux corresponding to a similar heliocentric distance, when the Crab is not occulted by a streamer). 
This could be because the line of sight to the Crab nebula passes through more coronal plasma during instances of streamer
occultation, leading to enhanced scatter broadening. In turn, this leads to a larger scatter-broadened image and a consequent reduction in the peak flux density.

\section{Turbulent density fluctuations and solar wind proton heating rate}

The angular broadening observations of the Crab nebula described in the previous section can be used to infer the 
amplitude of turbulent density fluctuations and associated heating rate of protons in the solar wind. The main quantity inferred from the 
observations is the structure function, which is essentially the spatial Fourier transform of the visibility 
observed with a given baseline. The structure function is used to estimate $C_{N}^2$, the so-called ``amplitude'' 
of the turbulent density spectrum. The density spectrum is modelled as a power law with an exponential cutoff at an 
``inner scale''. We assume that the inner scale is given by the proton inertial length. We elaborate on these aspects 
in the subsections below.

\subsection{Background electron density and the inner scale}\label{sec:dens}

Since our aim is to estimate the level of turbulent density fluctuations in relation to the background density ($N_e$), 
we use Leblanc density model \citep{Leb1998} to estimate the $N_e$ in the solar wind, 
\begin{equation}
N_e(R) = 7.2~R^{-2} + 1.95 \times 10^{-3}~R^{-4} + 8.1\times 10^{-7}~R^{-6} \,\,\,\, {\rm cm}^{-3}.
\label{leblanc}
\end{equation}
where `R' is the heliocentric distance in units of astronomical units (AU, 1 AU = $215 R_{\odot}$). 
The background electron density is used to compute the inner scale of the turbulent density spectrum. We assume that the inner 
scale $l_{i}$ is given by the proton inertial length \citep{Ver96,Lea99,Lea00,Smi01,Che14,Bru14}, which is related to
the background electron density by
\begin{equation}\label{eq:inner}
l_i(R) = v_A(R) / \Omega_p(R) = 2\pi/k_i(R)=228 \times \sqrt{N_{e}(R)}\, \, \, {\rm km},
\end{equation}

where $N_{e}$ is the electron density in ${\rm cm}^{-3}$, $k_i$ is the wavenumber, $v_A$ is the $\rm Alfv\acute{e}n$ speed 
and $\Omega_i$ is the proton gyrofrequency. 
We note that our definition differs slightly from that of \citet{Col89, Har89, Yam98} who use $l_i = 3 \times v_A(R)/\Omega_p(R)$ and $k_i = 3/l_i$.

\subsection{The structure function $D_{\phi}$}

The structure function $D_{\phi}(s)$ is defined by \citep{Pro75,Ish78,Col89,Arm90},

\begin{equation}
D_{\phi}(s)=-2 ln \Gamma(s)=-2ln\left[V(s)/V(0)\right] \, ,
\label{eq:struct}
\end{equation}

where the quantity $s$ represents the baseline length, $\Gamma(s)$ is the mutual coherence function, $V(s)$ denotes the visibility 
obtained with a baseline of length $s$ and $V(0)$ denotes the ``zero-length'' baseline visibility. The quantity $V(0)$ is the peak
flux density when the Crab nebula is situated far away from the Sun, and is unresolved; 
we set it to be $\approx 2015$ Jy at 80 MHz \citep{Bra1970, Mcl1985}. The images of the Crab nebula in 
Figure \ref{fig:graph_images} are obtained by combining the visibilities 
from all the baselines available in the GRAPH. We are interested in 
the turbulent density fluctuations at the inner scale, which is the scale at which the turbulent spectrum transitions from a power 
law to an exponential turnover. This is typically the smallest measurable scale; we therefore compute the structure function corresponding
to the longest available baseline (s = 2.6 km), since that corresponds to the smallest scale. 

\subsection{The amplitude of density turbulence spectrum ($C_N^2$)}\label{lab:cn2}

The turbulent density inhomogeneities are represented by a spatial power spectrum, comprising a power law together with an exponential turnover at the inner scale:

\begin{eqnarray}\label{eq:ss}
P_{\delta n}(k, R) = C_{N}^{2}(R) (\rho^2 ~k_x^2+k_y^2)^{-\alpha/2} \times \exp\biggl[ -(\rho^2 ~k_x^2+k_y^2)\bigg({l_{i}(R) \over 2 \pi }\bigg)^{2} \biggr]\, ,
\end{eqnarray}

where $k = \sqrt{\rho~ k_x^2+k_y^2}$ is the wavenumber, $k_x$ and $k_y$ are the wavenumber along and perpendicular to the 
large-scale magnetic field respectively. 
The quantity $\rho$ is a measure of the anisotropy of the turbulent eddies. In our calculations, we use the axial ratio of the scatter broadened images at 80 MHz (shown in Table \ref{tab:table-1}) for $\rho$.
The quantity $C_{N}^{2}$ 
is the amplitude of density turbulence, and has dimensions of ${\rm cm}^{-\alpha - 3}$, where $\alpha$ is the power law index of the 
density turbulent spectrum. At large scales the density spectrum follows the Kolmogorov scaling law with $\alpha=11/3$. 
At small scales, (close to the inner scale, when $s \approx l_i$) the spectrum flattens to $\alpha=3$ \citep{Col89}. 
Since we are interested in the density fluctuations near the inner scale, we use $\alpha = 3$. 

Many authors use analytical expressions for the structure function that are applicable in the asymptotic limits
$s \ll l_i$ or $s \gg l_i$ \citep{Col87, Arm00, Bas94, Pra11}. However, these expressions are not valid for situations 
(such as the one we are dealing with in this paper) where the baseline is comparable to the inner scale; i.e., $s \approx l_{i}$.
We therefore choose to use the General Structure Function (GSF) which is valid in the $s \ll l_i$ and $s \gg l_i$ regimes 
as well as when $s \approx l_i$ \citep{Ing2015a}. In the present case, largest baseline length $\approx 2.6$ km is comparable to the inner scale 
lengths $\approx 4.56$ km. The GSF is given by the following expression:

{\begin{eqnarray}
\label{eq:gsf}
\nonumber
{D_\phi(s)} = \frac{8 \pi^2 r_e^2 \lambda^2 \Delta L}{\rho~ 2^{\alpha-2}(\alpha-2)} {\Gamma \bigg( 1 - {{\alpha-2} \over 2} \bigg)}
	    {{C_N^2 (R) l_i^{\alpha-2}(R)} \over {(1 - f_p^2 (R) / f^2)}} \\
	     {\times \bigg\{ { _1F_1} {\bigg[ - {{\alpha-2} \over 2},~1,~ - \bigg( {s \over l_i(R)} \bigg)^2 \bigg]} -1 \bigg\}} \, \, \, \, {\rm rad}^{2},
\end{eqnarray}}
where ${ _1F_1}$ is the confluent hyper-geometric function, $r_e$ is the classical electron radius, 
$\lambda$ is the observing wavelength, $R$ is the heliocentric distance (in units of $R_{\odot}$), $\Delta L$ is the thickness of the scattering medium,
$f_p$ and f are the plasma and observing frequencies respectively. Substituting the model densities and $\alpha=3$ in 
Equation \ref{eq:gsf} enables us to calculate $C_N^2$. Following \citet{Sas2016}, we assume the thickness of the scattering screen to be
$\Delta L = (\pi/2) R_0$, where, $R_0$ is the impact parameter related to the projected heliocentric distance of the Crab nebula in units of cm. When the Crab nebula is occulted by a streamer, 
however, this estimate of $\Delta L$ is not valid. It is well known that the streamer owes its appearance to the fact that the line of sight to the streamer
intercepts excess coronal plasma that is contained around the current sheet ``fold''. It therefore stands to reason that the $\Delta L$ along a line of sight
that intercepts a streamer will be larger than that along a line of sight that does not include a streamer. In view of this, we use the formula 
$\Delta L = (\pi/2) R_0$ and compute the density fluctuation amplitude and turbulent heating rate only for the instances where the Crab nebula is not occulted by a streamer.

In the instances where it is occulted by a streamer, we can estimate the extra line of sight path length implied by 
the presence of the streamer. In order to do this, we first compute the structure function (Eq \ref{eq:gsf}) in 
the instances when the line of sight to the Crab nebula contains a streamer. We then estimate the ratio of this quantity 
to the structure function (at a similar heliocentric distance) when the line of sight does not intercept a streamer turns out to be $\approx 2$. 
For instance, $D_{\phi}(s = 2.6 \, {\rm km}, \,\, {\rm June}\, 17 \,2016)/D_{\phi}(s = 2.6 \,{\rm km}, \,\, {\rm June}\, 12 \,2016) = 2.16$.
On June 12 2016, the Crab nebula was situated at $10.18 R_{\odot}$ and the line of sight to it did not pass through a streamer. 
On June 17 2016, the Crab nebula was situated at a similar projected heliocentric distance ($10.2 R_{\odot}$),
but the line of sight to it passed through a coronal streamer. From Eq~(\ref{eq:gsf}), it is evident that this 
ratio is equal to the ratio of the $\Delta L$s in the two instances. In other words, the presence of a streamer approximately 
doubles the path length along the line of sight over which scattering takes place.

Although we show 80 MHz observations in this paper, we also have simultaneous observations at 53 MHz.
The structure function (equation \ref{eq:gsf}) is proportional to the square of the observing frequency 
(i.e., $D_{\phi}(s)~\propto~\lambda^2$). This predicts that the ratio of the structure functions at 80 and 53.3 MHz 
should be 0.44. Our observations yield a value of 0.43 for this ratio, and are thus consistent with the expected scaling.

\subsection{Estimating the density modulation index ($\epsilon_{N_e}=\delta N_{k_i} / N_e$)}\label{lab:densmod}

The density fluctuations $\delta N_{k_i}$ at the inner scale can be related to the spatial power spectrum (Equation \ref{eq:ss}) 
using the following prescription \citep{Cha2009}

\begin{equation}\label{eq:deltn}
{\delta}N_{k_i}^2(R) \sim 4 \pi k_i^3 P_{\delta N} (R, k_i) = 4 \pi C_{N}^{2}(R) k_i^{3 - \alpha} e^{-1} \,,
\end{equation}
where $k_{i} \equiv 2 \pi/l_{i}$. We estimate ${\delta}N_{k_i}$ by substituting $C_N^2$ calculated in \S~3.3 and 
using $\alpha = 3$ in Equation \ref{eq:deltn}. We then use this ${\delta}N_{k_i}$ and the background electron density 
($N_{e}$, \S~3.1) to estimate the density modulation index ($\epsilon_{N_e}$) defined by

\begin{equation}\label{eq:df}
\epsilon_{N_e}(R) \equiv {~\delta N_{k_{i}}(R) \over N_{e}(R)} \, 
\end{equation}

The density modulation index in the solar wind at different heliocentric distances is computed using Eq \ref{eq:df}. 
The results are listed in column (6) of table \ref{tab:table-1}.
The numbers in table \ref{tab:table-1} show that the density modulation index ($\epsilon_{N_e}$) in the solar wind
ranges from 1.9 $\times 10^{-3}$ to 7.7 $\times 10^{-3}$ in the heliocentric range $\approx 10-20~ R_{\odot}$. 
We have carried out these calculations only for the instances where the Crab nebula is not occulted by a streamer.

\subsection{Solar wind heating rate}\label{sec:hr}
We next use our estimates of the turbulent density fluctuations (${\delta}N_{k_i}$) to calculate the rate at 
which energy is deposited in solar wind protons, following the treatment of \citet{Ing2015b}. The basic assumption 
used is that the density fluctuations at small scales are manifestations of low frequency, 
oblique ($k_{\perp} \gg k_{\parallel}$), $\rm Alfv\acute{e}n$ wave turbulence. The quantities $k_{\perp}$ 
and $k_{\parallel}$ refer to components of the wave vector perpendicular and parallel to the background large-scale magnetic field respectively.
The turbulent $\rm Alfv\acute{e}n$ wave cascade transitions to such oblique $\rm Alfv\acute{e}n$ waves 
(often referred to as kinetic $\rm Alfv\acute{e}n$ waves) near the inner/dissipation scale. We envisage a 
situation where the turbulent $\rm Alfv\acute{e}n$ wave cascade resonantly damps on (and thereby heats) the 
protons at the inner scale. Since this implicitly assumes that the $\rm Alfv\acute{e}n$ waves do not couple 
to other modes at the inner scale, our estimate of the proton heating rate is an upper limit. As explained in 
\S~\ref{sec:dens}, we assume that the inner scale is the proton inertial length, which is expressible as $l_{i} = v_{\rm A}/\Omega_{p}$, 
where $v_{\rm A}$ is the $\rm Alfv\acute{e}n$ speed and $\Omega_{p}$ is the proton gyrofrequency. This way of writing the the proton inertial 
length emphasizes its relation to the resonant damping of $\rm Alfv\acute{e}n$ waves on protons. 

The specific energy per unit time ($\epsilon\, , \, {\rm erg ~cm^{-3}~s^{-1}}$) in the turbulent $\rm Alfv\acute{e}n$ wave cascade is transferred from 
large scales to smaller ones, until it dissipates at the inner/dissipation scale. The proton heating rate equals the turbulent energy cascade rate 
at the inner scale ($\epsilon_{k_i}$), which is given by \citep{Hol1999, Cha2009, Ing2015b}, 
\begin{equation}\label{eq:hr}
\epsilon_{k_i}(R)=c_0 \rho_p k_i(R) \delta v_{k_i}^3(R) ~ \rm erg ~cm^{-3}~s^{-1} \, ,
\end{equation}

where $c_0$ is a constant usually taken to be 0.25 \citep{How2008, Cha2009} and $\rho_p=m_pN_e(R)~\rm g~ cm^{-3}$, 
with $m_p$ representing the proton mass in grams. The quantity $k_i=2 \pi/l_i$ is the wavenumber corresponding to the
inner scale (Eq~\ref{eq:inner}) and $\delta v_{k_i}$ represents the magnitude of turbulent velocity fluctuations at
the inner scale. The density modulation index $\epsilon_{N_e}$ and the turbulent velocity fluctuations are related via 
the kinetic $\rm Alfv\acute{e}n$ wave dispersion relation \citep{How2008,Cha2009,Ing2015b}


\begin{eqnarray}\label{eq:rmsv}
 \delta v_{k_i}(R)=\Bigg({1+{\gamma_i k_i^2(R) \rho_i^2(R)} \over {k_i(R) l_i(R)}} \Bigg) \epsilon_{N_e} (R, k_i) v_A(R) \, .
\end{eqnarray}

The adiabatic index $\gamma_i$ is taken to be 1 \citep{Cha2009} and the proton gyroradius ($\rho_i$) is given by

\begin{equation}\label{eq:rho_i}
\rho_i(R)=102 \times \mu^{1/2} T_i^{1/2} B^{-1}(R)~\rm cm,
\end{equation}

where $\mu$ is the ion mass expressed in terms of proton mass ($\approx 1$) and $T_i$ is the proton temperature in eV. We use $T_i=86.22$ eV which corresponds to a temperature of $1 \times 10^6$ K.

\begin{figure}[!ht]
\centerline{\includegraphics[width=16cm]{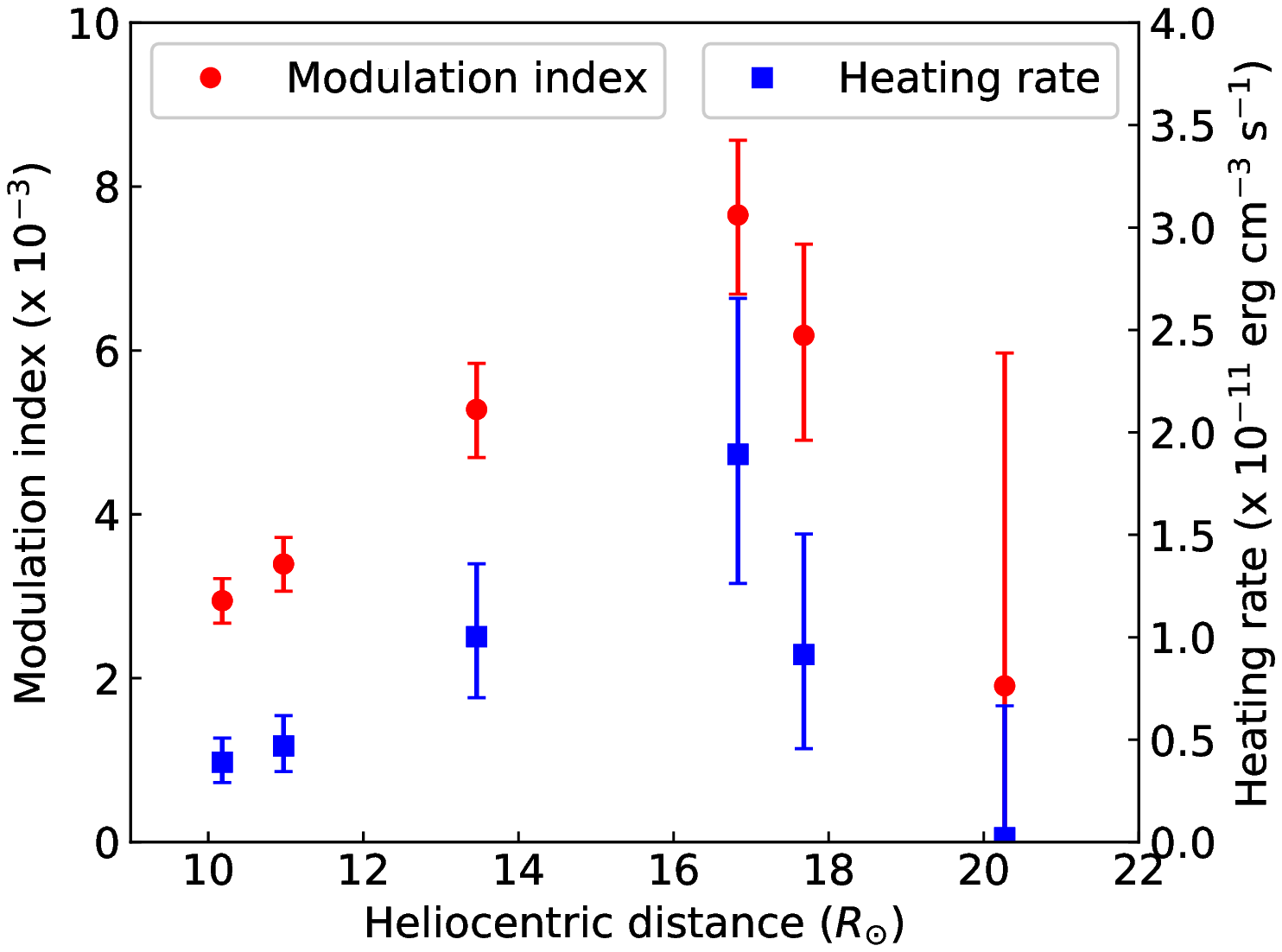}}
\caption{The variation of the density modulation index (red circles) and the solar wind 
proton heating rate (blue squares) with projected heliocentric distance. We note that the proton heating rate is 
correlated with the density modulation index.} 
\label{fig:hr}
\end{figure}

The $\rm Alfv\acute{e}n$ speed ($v_A$) in the solar wind is given by 

\begin{equation}\label{eq:va}
v_A(R)=2.18\times 10^{11} \mu^{-1/2} N_e^{-1/2}(R)B(R) ~\rm cm~s^{-1}, 
\end{equation}

and the magnetic field stength (B) is taken to be the Parker spiral mangetic field in the ecliptic 
plane \citep{Wil1995} 

\begin{equation}\label{eq:parker}
B(R)= 3.4 \times 10^{-5} R^{-2} (1+R^2)^{1/2} ~ \rm Gauss, 
\end{equation}

where, `R' is the heliocentric distance in units of AU. 
Equations (\ref{eq:parker}), (\ref{eq:va}), (\ref{eq:rho_i}), (\ref{eq:rmsv}) and 
the density modulation index computed in \S~\ref{lab:densmod} are used in Eq~(\ref{eq:hr}) 
to compute the solar wind heating rate at different heliocentric distances. These values are tabulated in column (7)
of Table \ref{tab:table-1}. Figure~\ref{fig:hr} depicts the density modulation index and the solar wind heating rate 
graphically as a function of heliocentric distance.

\begin{table}
\centering
\vspace*{5px}

 \begin{tabular}{|c|c |c| c| c| c| c| c| c| c| c| c| c| c| c| c| c|} 
 \hline
 \hline
 S.No & Date & R & Peak flux density& $\rm \rho $ & $\epsilon_{N_e}$ & Heating rate \\
   &  & $\rm (R_{\odot})$ & (Jy) & & & ($\rm erg~ cm^{-3}~ s^{-1}$) \\
 (1)  & (2) & (3) & (4) & (5) & (6) & (7) \\
  
   \cline{1-7}
    \multicolumn{7}{|c|}{Line of sight to the Crab does not include a streamer} \\
        \cline{1-7}

1 &	12 June 2016 &	10.18 &	1349 &	1.48 &	 2.9E-3 &	 3.9E-12\\
2 &	18 June 2016 &	13.46 &	1473 &	1.76 &	 5.3E-3 &	 1.0E-11\\
3 &	19 June 2016 &	16.83 &	1546 &	1.69 &	 7.7E-3 &	 1.9E-11\\
4 &	20 June 2016 &	20.27 &	2003 &	1.98 &	 1.9E-3 &	 2.2E-13\\
5 &	09 June 2017 &	21.13 &	2015 &	1.48 &	 - 	  &		- \\
6 &	10 June 2017 &	17.68 &	1732 &	1.57 &	 6.2E-3 &	 9.2E-12\\
7 &	12 June 2017 &	10.97 &	1386 &	1.50 &	 3.4E-3 &	 4.7E-12\\
8 &	22 June 2017 &	26.34 &	2015 &	1.40 &	- 	  & 		- \\

\cline{1-7} 
    \multicolumn{7}{|c|}{Line of sight to the Crab includes a streamer} \\
        \cline{1-7}						
9 &	17 June 2016 &	10.20 &	845  &	2.44 &	- & -\\
10 &	17 June 2017 &	9.41  &	901  &	2.51 &	 - & - \\
11 &	18 June 2017 &	12.61 &	800  &	1.65 &	 - & - \\

\hline
\hline
\end{tabular}
\caption{The table describes the observational quantities and the derived plasma parameters in the solar wind. 
}
\label{tab:table-1}

\end{table}

\section{Summary and conclusions}
\subsection{Summary}
We have imaged (figure~\ref{fig:graph_images}) the Crab nebula at 80 MHz using the GRAPH in June 2016 and 2017, when it passed close to the Sun, and was obscured by the turbulent solar wind. 
Since the Crab nebula is a point source at 
80 MHz when it is far from the Sun, these images are evidence of anisotropic 
scatter-broadening of radiation emanating from it as it passes through the 
turbulent solar wind. We calculate the structure function 
with the visibilities from the longest baselines (2.6 km) used in making these images. 
The structure function is used to infer the amplitude of the density turbulence spectrum ($C_{N}^{2}$), which is then used to 
compute the magnitude of the turbulent density fluctuations at the inner scale (Eq~\ref{eq:deltn}). This is then used to compute 
the density modulation index (Eq~\ref{eq:df}). Assuming that the turbulent $\rm Alfv\acute{e}n$ wave cascade in the solar wind dissipates on 
protons at the inner scale, we calculate the heating rate of protons in the solar wind (Eq~\ref{eq:hr}). The density modulation index and 
solar wind proton heating rate are plotted in Figure~\ref{fig:hr} as a function of heliocentric distance.


\subsection{Conclusions}
The main conclusions of this paper pertain to the anisotropy of the scatter-broadened image of the Crab nebula, 
the density modulation index of the turbulent fluctuations in the solar wind and the solar wind proton heating
rate from $9-20~R_{\odot}$. Some of the conclusions are:

\begin{itemize}

\item
The 80 MHz scatter broadened images of the Crab nebula at heliocentric distances ranging from $9$ to $20~R_{\odot}$ 
in the solar wind are anisotropic, with axial ratios typically $\lesssim 2$ (table~\ref{tab:table-1}). The major axis 
of the Crab nebula is typically oriented perpendicular to the magnetic field direction, as in \citet{Ana94,Arm90} 
(although their observations were at much smaller distances from the Sun).



\item 

On 17 June 2016 and 17 June 2017, a coronal streamer was present along the line of sight to the Crab nebula. 
The line of sight to the Crab encountered more coronal plasma on these days, as compared to the days when a streamer 
was not present. The axial ratio of the scatter-broadened images on these days was somewhat larger ($\approx 2$, 
see table~\ref{tab:table-1}) and the peak flux density is considerably lower (figure~\ref{fig:peak}), reflecting 
this fact. In the presence of a streamer, the path length over which scattering takes place was found to be approximately 
twice of that when the streamer was not present. 


\item

The density modulation index ($\epsilon_{N_{e}} \equiv \delta N_{e}/N_{e}$) at the inner scale of the turbulent spectrum 
in the solar wind from $9-20~R_{\odot}$ ranges from 1.9 $\times 10^{-3}$ to 7.7 $\times 10^{-3}$ (see table~\ref{tab:table-1}). 
Earlier estimates of $\epsilon_{N_e}$ include \citet{Sas2016} who reported 
$0.001 \lesssim \epsilon_{N_e} \lesssim 0.1$ from 10-45 $R_{\odot}$, $0.001 \lesssim \epsilon_{N_e} \lesssim 0.02$ reported by 
\citet{Bis14b} in the distance range 56-—185 $R_{\odot}$ and $0.03 \lesssim \epsilon_{N_e} \lesssim 0.08$ reported by 
\citet{Spa04} at 1 AU (215 $R_{\odot}$). The red circles in Figure \ref{fig:hr} depict the modulation index as a function of
heliocentric distance. Figure \ref{fig:hr} shows that the modulation index in the heliocentric distance $12-18~R_{\odot}$ is 
relatively higher. As explained in \citet{Sas2016}, this might be because the line of sight to the Crab nebula at these distances
passes through the fast solar wind, which has relatively higher proton temperatures \citep{Lop1986}. Furthermore, the density 
modulation index is correlated with the proton temperature \citep{Cel87}. Taken together, this implies that one could expect 
higher values for the density modulation index in the fast solar wind.
 

\item We interpret the turbulent density fluctuations as manifestations of kinetic $\rm Alfv\acute{e}n$ wave turbulence at
small scales. Assuming that the turbulent $\rm Alfv\acute{e}n$ wave cascade damps resonantly on the protons at the inner scale,
we use our estimates of the density modulation index to calculate the proton heating rate in the solar wind. We find that the
estimated proton heating rate in the solar wind from $9-20~R_{\odot}$ ranges from 
$2.2 \times 10^{-13}$ to $1.0 \times 10^{-11} ~\rm erg~cm^{-3}~s^{-1}$ (blue squares in figure~\ref{fig:hr}).

\end{itemize}

\section{Acknowledgments}
KSR acknowledges the financial support from the Science $\&$ Engineering Research Board (SERB), Department of Science $\&$ Technology, India 
(PDF/2015/000393). PS acknowledges support from the ISRO RESPOND program. AV is supported by NRL grant N00173-16-1-G029. We thank the staff of the Gauribidanur
observatory for their help with the observations and maintenance
of the antenna and receiver systems there. KSR acknowledges C. Kathiravan for the valuable discussions related to the GRAPH observations. 
SOHO/LASCO data used here are produced by a consortium of the Naval 
Research Laboratory (USA), Max-Planck-Institut fuer 
Aeronomie (Germany), Laboratoire d'Astronomie (France), and the University of Birmingham (UK). 
SOHO is a project of international cooperation between ESA and NASA.
The authors would like to thank the anonymous referee for the valuable and constructive suggestions.
\bibliographystyle{apj}
\bibliography{ms}

\end{document}